\documentclass[%
 reprint,
superscriptaddress,
 amsmath,amssymb,
 aps,
]{revtex4-2}

\usepackage{graphicx}
\usepackage{dcolumn}
\usepackage{bm}

\usepackage{physics}
\usepackage{upgreek}
\usepackage{gensymb}
\usepackage{units}
\usepackage[markup=underlined]{changes}
\usepackage[colorlinks=true,linkcolor=blue,urlcolor=blue,citecolor=blue]{hyperref}

\begin{document}


\title{Quantum interference between quasi-2D Fermi surface sheets in UTe$_2$}

\author{T. I. Weinberger}
\thanks{These authors contributed equally to this work.}
\author{Z. Wu}
\thanks{These authors contributed equally to this work.}
\affiliation{Cavendish Laboratory, University of Cambridge,\\
 JJ Thomson Avenue, Cambridge, CB3 0HE, United Kingdom}

\author{D.~E.~Graf}
\affiliation{National High Magnetic Field Laboratory, Tallahassee, Florida, 32310, USA}

\author{Y. Skourski}
\affiliation{Hochfeld-Magnetlabor Dresden (HLD-EMFL),\\ Helmholtz-Zentrum Dresden-Rossendorf, Dresden, 01328, Germany}

 \author{A. Cabala}
 \author{J. Pospíšil} 
 \author{J. Prokleška} 
 \author{T.~Haidamak} 
 \author{G. Bastien}
 \author{V. Sechovský}
 \affiliation{Charles University, Faculty of Mathematics and Physics,\\ Department of
Condensed Matter Physics, Ke Karlovu 5, Prague 2, 121 16, Czech Republic}

\author{G. G. Lonzarich}
\affiliation{Cavendish Laboratory, University of Cambridge,\\
 JJ Thomson Avenue, Cambridge, CB3 0HE, United Kingdom}
 
 \author{M. Vali{\v{s}}ka}
 \affiliation{Charles University, Faculty of Mathematics and Physics,\\ Department of
Condensed Matter Physics, Ke Karlovu 5, Prague 2, 121 16, Czech Republic}

\author{F. M. Grosche}
\author{A. G. Eaton}
 \email{alex.eaton@phy.cam.ac.uk}
\affiliation{Cavendish Laboratory, University of Cambridge,\\
 JJ Thomson Avenue, Cambridge, CB3 0HE, United Kingdom}
 
\date{\today}

\begin{abstract}
\noindent
UTe$_2$ is a spin-triplet superconductor candidate for which high quality samples with long mean free paths have recently become available, enabling quantum oscillation measurements to probe its Fermi surface and effective carrier masses. It has recently been reported that UTe$_2$ possesses a 3D Fermi surface component [\href{https://doi.org/10.1103/PhysRevLett.131.036501}{Phys. Rev. Lett. \textbf{131}, 036501 (2023)}]. The distinction between 2D and 3D Fermi surface sections in triplet superconductors can have important implications regarding the topological properties of the superconductivity. Here we report the observation of oscillatory components in the magnetoconductance of UTe$_2$ at high magnetic fields. We find that these oscillations are well described by quantum interference between quasiparticles traversing semiclassical trajectories spanning magnetic breakdown networks. Our observations are consistent with a quasi-2D model of this material's Fermi surface based on prior dHvA-effect measurements. Our results strongly indicate that UTe$_2$ -- which exhibits a multitude of complex physical phenomena -- possesses a remarkably simple Fermi surface consisting exclusively of two quasi-2D cylindrical sections.

\end{abstract}

\maketitle

Young's double slit experiment represents a powerful example of the wave-particle duality of photons~\cite{young1804}. A century later Davisson and Germer observed a similar phenomenon involving the quantum mechanical interference of a beam of electrons incident on a crystalline target~\cite{davisson1927nature,DavissonPhysRev.30.705}. In the solid state, superconducting quantum interference devices provide exceptionally accurate measurements of magnetic flux via diffraction-modulated interferometry~\cite{josephson1962,Mercereau-superconductivity-book}. For the case of normal metals, the manifestation of quantum interference (QI) effects in the magnetoconductance was first predicted by Shiba and Fukuyama~\cite{shiba1969}, and soon thereafter experimentally realized by Stark and Friedberg in their measurements of the magnetoresistance of magnesium~\cite{stark1971quantum}. The concept of the Stark interferometer is based on interference between semiclassical quasiparticle trajectories across magnetic breakdown networks connecting separate Fermi surface (FS) sections, yielding oscillations in the conductivity that are periodic in inverse magnetic field strength~\cite{stark1974interfering,stark1977quantitative,morrison_stark_1981two,Shoenberg1984}.

Since the seminal experiments by Stark and coworkers, quantum interference oscillations (QIOs) have been observed in a variety of materials~\cite{UjiPhysRevB.53.14399,HarrisonLaB6PhysRevLett.80.4498,HarrisonCeB6PhysRevLett.81.870,GrafPhysRevB.75.245101,nairPhysRevB.102.075402} including, in particular, a number of organic metals with quasi-2D (Q2D) FSs \cite{harrison1996,KartsovnikPhysRevLett.77.2530,singletonrepprogphys2000,schrama2001,ProustPhysRevB.65.155106,Kartsovnik_review,audouard2004,lyubovskii2019}. Unlike quantum oscillations (QOs) from the dHvA- or SdH-effects, in which phase coherence and Landau quantization of quasiparticles traversing orbits corresponding to closed FS sections provide a direct measurement of the FS~\cite{Shoenberg1984}, QIOs only yield an indirect probe of the FS, as their frequencies correspond to $k$-space orbits spanning separate FS sections. Therefore, QIOs are only observed in materials in which the $k$-space separation of FS sections is sufficiently small for quasiparticles to tunnel between FS sheets in accessible magnetic field strengths~\cite{Chambers_1966,Shoenberg1984}. It is important to note that QI is exclusively a kinetic effect and is thus observable in the electrical transport -- unlike the dHvA-effect, QIOs do not correspond to an oscillatory component of the free energy, therefore QI effects \textit{cannot} be observed in bulk thermodynamic properties such as the magnetization~\cite{stark1977quantitative,morrison_stark_1981two,harrison1996,kaganov1983coherent}.

Here, we report the observation of QIOs at high magnetic fields in contactless resistivity measurements of the heavy fermion paramagnetic metal UTe$_2$. This material has recently shown promising signs of being a spin-triplet superconductor~\cite{Ran2019Science,Ranfieldboostednatphys2019,Aoki_UTe2review2022}, similar to the analogous ferromagnetic compounds UGe$_2$, URhGe and UCoGe~\cite{Montu_UGe2,URhGE_Aoki2001,UCoGe_PhysRevLett.99.067006}. Evidence indicating triplet pairing in UTe$_2$ comes from a number of sources including small changes in the NMR Knight shift on cooling through the superconducting critical temperature $T_\text{c}$~\cite{Aoki_UTe2review2022,aoki_AuNMR_2023} along with anisotropic upper critical fields that far exceed the Pauli limit for singlet pairing~\cite{chandrasekhar1962note,Clogston_PhysRevLett.9.266,Ran2019Science,tony2023enhanced}. Recent advances in the growth procedure of single crystal UTe$_2$ specimens have led to a marked enhancement in crystalline quality, enabling the observation of QOs from the dHvA-effect~\cite{AokidHvA_UTe2-2022,Eaton2024}. The angular profile of the dHvA data is indicative of a relatively simple Q2D FS, consisting of one electron-type and one hole-type cylinder, each hosting quasiparticles of heavy effective masses $\sim$~40~$m_{\text{e}}$~\cite{AokidHvA_UTe2-2022,Eaton2024}.

\textit{Methods} -- UTe$_2$ single crystals were grown by a molten salt flux technique~\cite{PhysRevMaterials.6.073401MSF_UTe2} using the methodology detailed in ref.~\cite{Eaton2024}. This technique has been shown to yield high quality specimens of $T_{\text{c}} \approx$~2.1~K with long mean free paths of the order of 100~nm~\cite{tony2023enhanced,AokidHvA_UTe2-2022,PhysRevMaterials.6.073401MSF_UTe2,Eaton2024}. Details regarding sample characterization are given in the Supplemental Materials~\footnote{See \href{https://journals.aps.org/prl/supplemental/10.1103/PhysRevLett.132.266503/UTe2_QIOs_final_SI.pdf}{Supplemental Material} for additional data and discussion, which
includes Refs.~\cite{semeniuk2023truncated,ramshaw2015science,indy2022npj,Rosa2022,BalakirevPhysRevB.91.220505,MSFmuonsRosaSonier23,U3Te5TOUGAIT1998,Aoki_magnetic-pressureUTe2-2021,onsager,Dingle,weinberger2024pressureenhanced}}. Contactless resistivity measurements were performed in static fields to 41.5~T at the National High Magnetic Field Lab, Tallahassee, Florida, using the tunnel diode oscillator (TDO) technique~\cite{RevSciIns_TDO}; similar measurements were obtained in pulsed fields to 70~T at the Hochfeld-Magnetlabor, HZDR, Dresden, using the proximity detector oscillator (PDO) technique~\cite{PDO_Altarawneh,ghannadzadeh2011PDO}.

\begin{figure*}[t!]
\includegraphics[width=\linewidth]{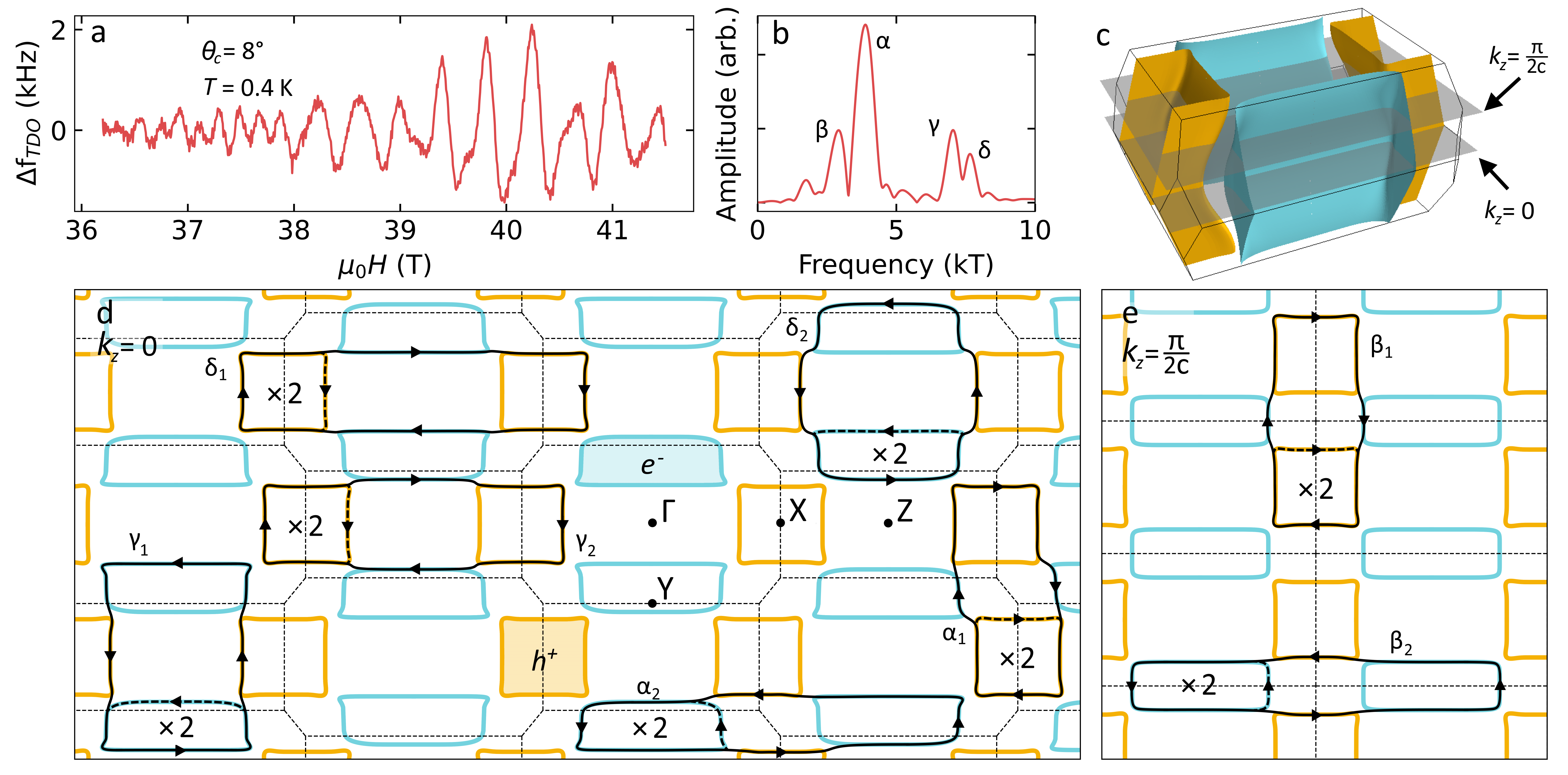}
\caption{\label{fig:wiggle1} (a) QIOs in the contactless resistivity of UTe$_2$ and (b) the corresponding FFT spectra. (c) Our Fermi surface model for UTe$_2$ adapted from ref.~\cite{Eaton2024}, with the planes $k_{z} = 0$ and $k_{z} = \frac{\pi}{2c}$ indicated. (d) An extended-zone view, with the $c$-axis into the page, at $k_{z} = 0$ and (e) at $k_{z} = \frac{\pi}{2c}$. QI trajectories enclosing areas that correspond to the QIO frequencies $\upalpha$-$\updelta$ are indicated; note that each enclosed area $\upalpha$-$\updelta$ has two distinct MB networks corresponding to it (denoted as $\lambda_{1,2}$).}
\end{figure*}

\textit{Results} -- Figure~\ref{fig:wiggle1} shows the background-subtracted TDO signal ($\Delta f_{\text{TDO}}$) for magnetic field oriented 8$\degree$ away from the crystalline $c$-axis towards the $a$-axis ($\theta_c = 8\degree$)\footnote{The datasets supporting the findings of this study are available from the University of Cambridge Apollo Repository [\href{https://doi.org/10.17863/CAM.108558}{doi.org/10.17863/CAM.108558}].}. The FFT of the $\Delta f_{\text{TDO}}$ data reveals four clear frequency branches, which we label as $\upalpha$-$\updelta$. Notably, the FFT spectra at $\theta_c = 8\degree$ of the TDO signal are very different to the spectra we observed in our prior dHvA study at the same angle (ref.~\cite{Eaton2024}), implying that these are not QOs stemming from the SdH-effect. Furthermore, the amplitude of dHvA QOs diminished by almost an order of magnitude between 19~mK and 200~mK -- whereas here the signal is large and very well resolved at 400~mK. These observations indicate that the oscillations in $f_{\text{TDO}}$ are likely QIOs not QOs, as QIOs generally correspond to reciprocal space areas constructed from sums and differences between FS sections, and often exhibit effective masses much lower than those of dHvA and SdH QOs~\cite{harrison1996,KartsovnikPhysRevLett.77.2530,HarrisonLaB6PhysRevLett.80.4498}.

\begin{figure}[h!]
\includegraphics[width=0.6\linewidth]{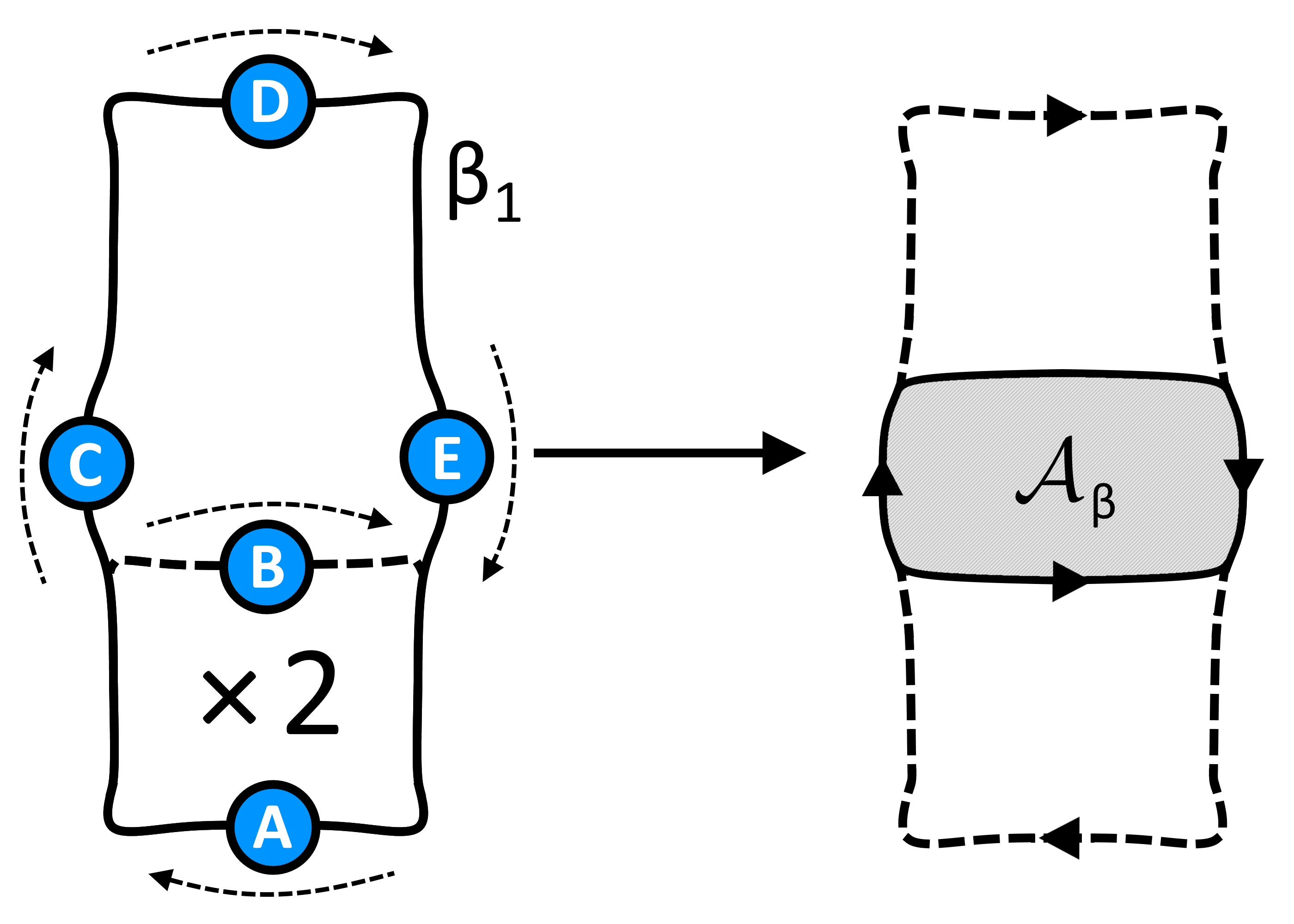}
\caption{\label{fig:wiggle2} Schematic of two semiclassical quasiparticle trajectories that interfere to give the $f_{\upbeta}$ frequency component. The difference in area of the paths ACDEA and ABABA is equal to $\mathcal{A}_{\upbeta}$, as shown in the text. Note that the orbit CDEBC is not possible due to the direction of the Lorentz force (indicated with arrows). The frequency components $\upalpha$, $\upgamma$ and $\updelta$ come from QI between analogous networks, as traced out in Figure~\ref{fig:wiggle1}.}
\end{figure}

Using our FS model from ref.~\cite{Eaton2024}, we illustrate in Figs.~\ref{fig:wiggle1} \& \ref{fig:wiggle2} how the frequencies of the $\upalpha$-$\updelta$ FFT peaks correspond remarkably well to $k$-space areas \textit{between} the cylindrical Fermi sheets, which are centred at the centre and corners of the first Brillouin zone (BZ). Each of these frequency components can thus be well understood as coming from QI between two quasiparticles -- one making two orbits around a FS cylinder, and the other traversing a magnetic breakdown (MB) network between two cylinders of the same carrier type.

\begin{figure}[h!]
\includegraphics[width=1\linewidth]{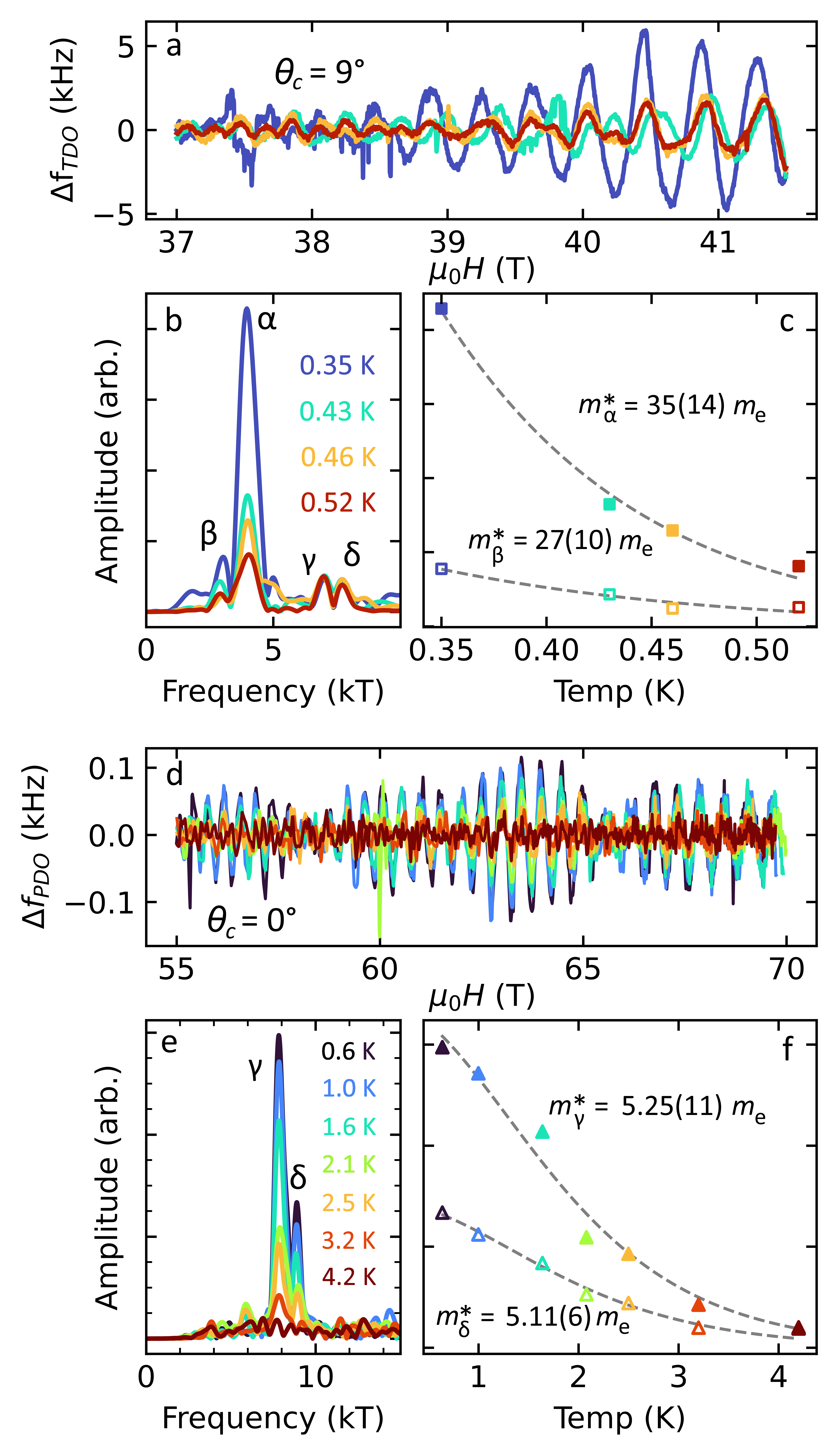}
\caption{\label{fig:wiggle3} QIOs, FFTs, and apparent effective masses for (a-c) steady field measurements focussing on the $\upalpha$ and $\upbeta$ frequency components, and (b-e) higher temperature pulsed field measurements focussing on the $\upgamma$ and $\updelta$ components. The effective masses $m^{*}_{\upgamma}$ and $m^{*}_{\updelta}$ are markedly lower than those observed in dHvA-effect measurements for the same field orientation~\cite{Eaton2024}.}
\end{figure}

\begin{figure}[h!]
\includegraphics[width=1\linewidth]{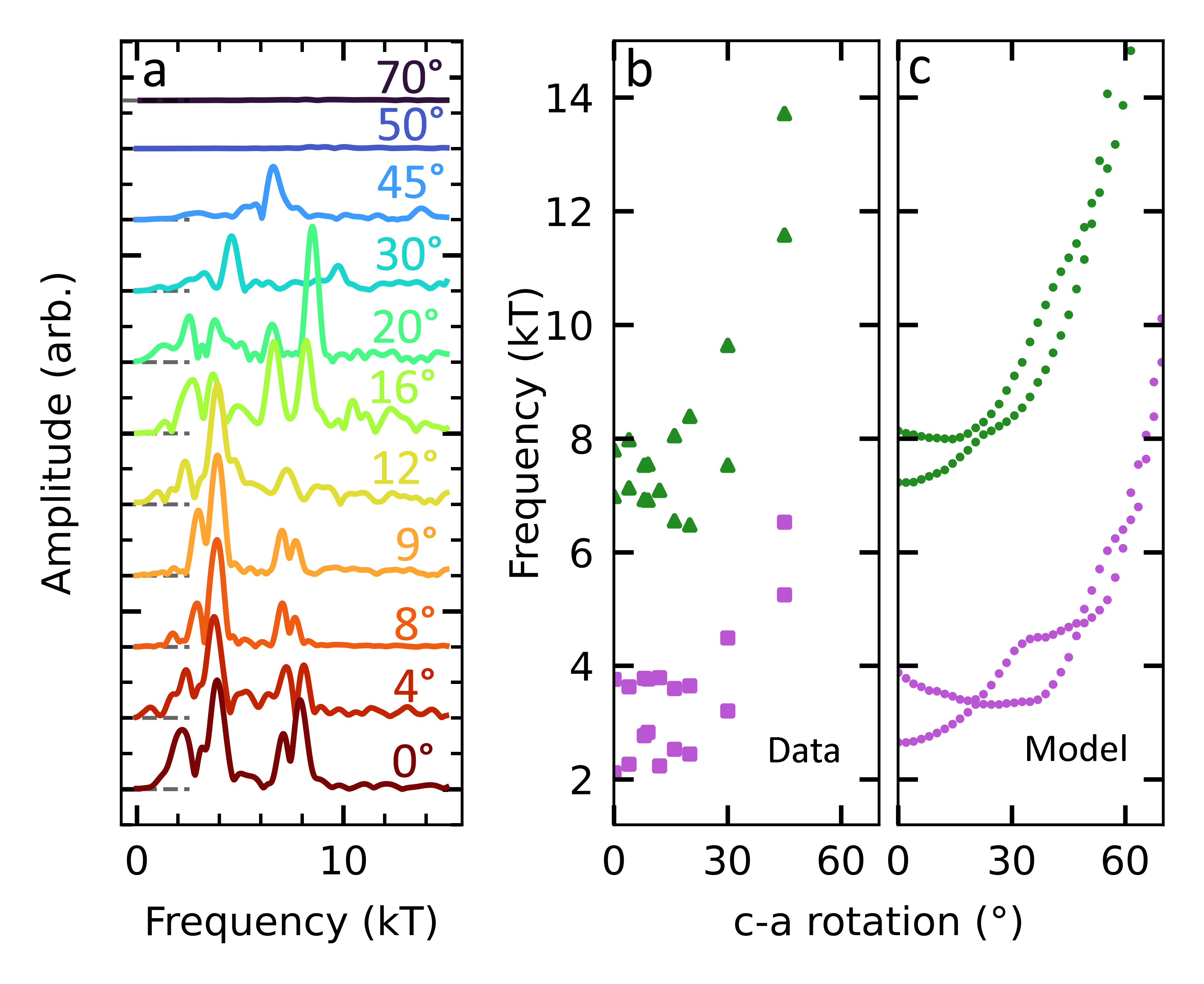}
\caption{\label{fig:wiggle4}(a) Evolution of FFT spectra as a function of magnetic field tilt angle away from the $c$-axis. (b) Frequency versus angle for the $\upalpha$ and $\upbeta$ branches (in purple) and the $\upgamma$ and $\updelta$ branches (in green). (c) The expectation of the angular frequency profile corresponding to the areas $\mathcal{A}_{\upalpha,\upbeta,\upgamma,\updelta}$ computed from our FS model depicted in Fig.~\ref{fig:wiggle1}. Surprisingly good correspondence between model prediction and experimental measurement is observed, given the simplicity of the model assumptions (given in the Supplementary Materials).}
\end{figure}

\begin{figure*}[t!]
\includegraphics[width=1\linewidth]{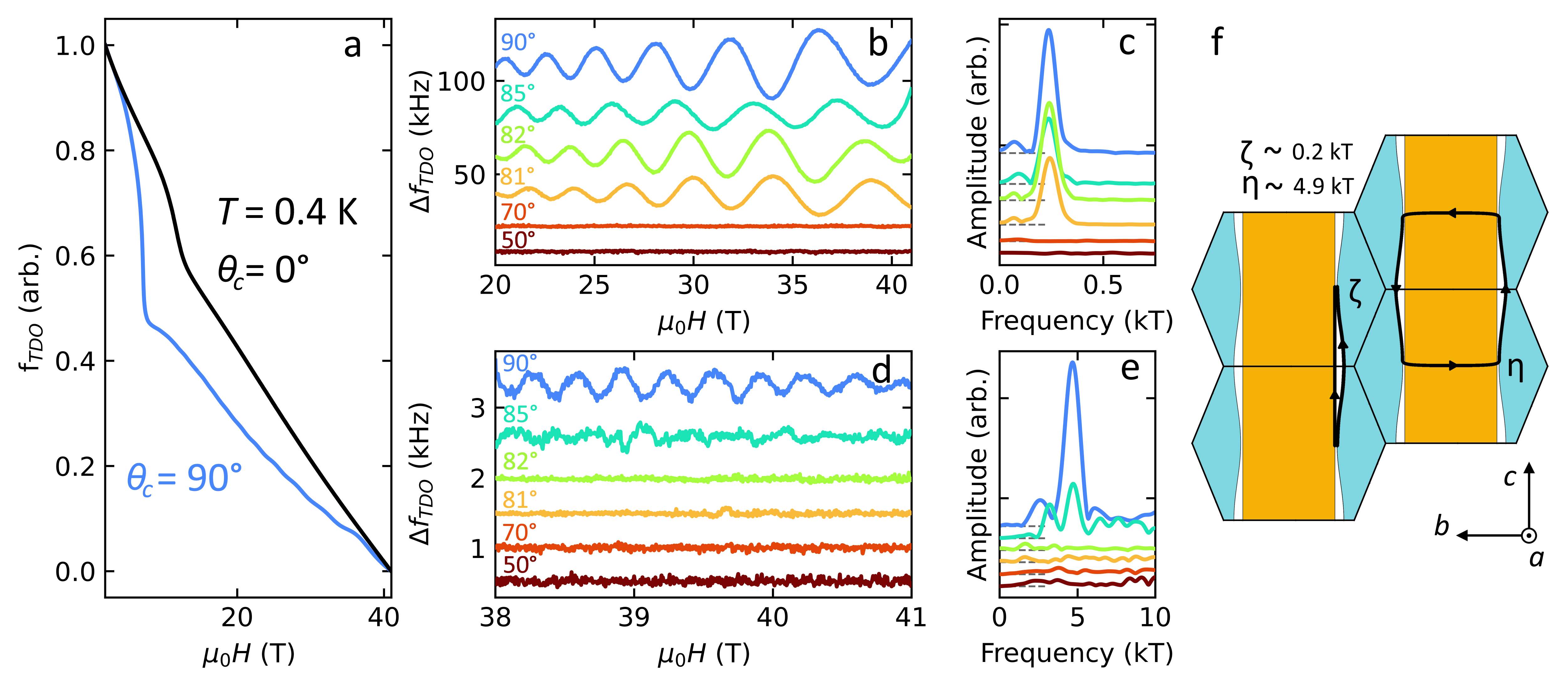}
\caption{\label{fig:wiggle5}(a) Raw TDO signal for magnetic field oriented along the $c$-axis ($\theta_c = 0\degree$) and along the $a$-axis ($\theta_c = 90\degree$). (b) QIOs at $\theta_c$ angles as indicated over 20-41.5~T and (c) their corresponding FFTs. (d) High frequency oscillations at $\theta_c = 90\degree$ are quickly suppressed by rotation away from the $a$-axis. (e) FFTs of the curves in panel d. Only the data at $\theta_{c} = 90\degree$ has a resolvable frequency component above the noise floor of the measurement. (f) A side-view of the cylindrical sheets of our Q2D FS model (compared to the axial-view given in Fig.~\ref{fig:wiggle1}). The trajectories $\upzeta$ and $\upeta$ are identified, which enclose areas with very good correspondence to the oscillatory frequencies observed for magnetic field orientated along the $a$-axis.}
\end{figure*}

To show this, we consider the generalized theory of MB orbits given by Kaganov and Slutskin~\cite{kaganov1983coherent}. In a magnetic field $B$ the oscillatory component of a kinetic coefficient, such as the electrical conductivity, is composed of harmonics of the form

\begin{equation}
    \sum_{\lambda,\lambda^{'}} \exp[i \left( \phi_{\lambda} - \phi_{\lambda^{'}} \right)] = \sum_{\lambda,\lambda^{'}} \exp\left( \frac{i \hbar }{eB} \mathcal{A}_{\lambda,\lambda^{'}}\right)
\label{eqn:sums}
\end{equation}

\noindent
for phase $\phi$ where $\lambda$ and $\lambda^{'}$ are the two semiclassical quasiparticle paths that share a common start and end point, enclosing between them an area in reciprocal space of $\mathcal{A}_{\lambda,\lambda^{'}}$~\cite{KartsovnikPhysRevLett.77.2530}\footnote{Note that the calculation of $\mathcal{A}_{\lambda,\lambda^{'}}$ is dependent on both the number and the direction of the trajectories included in $\lambda$ and $\lambda^{'}$}. 


Take for example the area $\mathcal{A_\upbeta}$ shaded in Fig.~\ref{fig:wiggle2}, which sits at the corner of the first BZ for $k_z = \frac{\pi}{2c}$. Writing the area of the hole-type FS cylinder as $\mathcal{A}_{h^{+}}$, we can see that the area $\mathcal{A_\upbeta}$ is equal to the difference of the areas enclosed by the paths $\lambda$ = ACDEA and $\lambda^{'}$ = ABABA as $\mathcal{A_\text{ACDEA}} - \mathcal{A}_\text{ABABA} = (2\mathcal{A}_{h^{+}} + \mathcal{A}_\upbeta) - 2\mathcal{A}_{h^{+}} = \mathcal{A_\upbeta}$. Similarly, areas corresponding to the $\upalpha$, $\upgamma$ and $\updelta$ frequency components are formed by QI between the quasiparticle trajectories traced in Fig.~\ref{fig:wiggle1}~\footnote{Each of $\mathcal{A}_{\lambda} = \{\mathcal{A}_{\upalpha}, \mathcal{A}_{\upbeta}, \mathcal{A}_{\upgamma}, \mathcal{A}_{\updelta}\}$ have two distinct QI paths corresponding to them, each requiring only 4 instances of MB, which we label as $\lambda_{1,2}$}. The probability of a quasiparticle traversing a path depends on the number of MB tunnelling events (each of probability amplitude $p$) and Bragg reflections (of probability amplitude $q$) that are contained within the path, where $\abs{p}^2 = P = \exp(\nicefrac{-B_{0}}{B})$ for breakdown field $B_{0}$ and $q = i\sqrt{\left( 1-P \right)}$~\cite{Shoenberg1984,Chambers_1966,harrison1996,falicovPhysRev.147.505,KartsovnikPhysRevLett.77.2530}. Therefore, the probability amplitudes for quasiparticles to traverse the paths $\lambda$ = ACDEA and $\lambda^{'}$ = ABABA, corresponding to the $\upbeta$ frequency in Fig.~\ref{fig:wiggle2}, are $q^{4} p^{4} \exp(i \phi_{\lambda})$ and $q^{8} \exp(i \phi_{\lambda^{'}})$, respectively. Due to this exponentially suppressed tunnelling probability -- which necessitates the application of high magnetic fields -- we limit our discussion just to the lowest order relevant networks as depicted in Fig.~\ref{fig:wiggle1}, each of which requires only 4 instances of MB.

By Eqn.~\ref{eqn:sums}, the probability of quasiparticles traversing the paths in Fig.~\ref{fig:wiggle2} will involve oscillating terms including some proportional to $\cos\left[{\phi_{\lambda}-\phi_{\lambda^{'}}}\right] = \cos\left[{2\pi (2f_{h^{+}} + f_{\upbeta} - 2f_{h^{+}})/B}\right] = \cos\left[{2\pi f_{\upbeta}/B}\right]$, which will contribute to the (real part of the) conductivity. Furthermore, in the low temperature limit the temperature dependence of QIOs simply follows the Lifshitz-Kosevich theory~\cite{Lifshitz96.963,Shoenberg1984,HarrisonLaB6PhysRevLett.80.4498} with an apparent effective mass $m^*_{\lambda,\lambda'}$, which is proportional to the dependence of the phase on the electron energy, $E_{k}$:


\begin{equation}
   m^*_{\lambda,\lambda'} = \frac{e \hbar B}{2\pi} \left|\frac{\partial\left(\phi_{\lambda}-\phi_{\lambda^{'}}\right)}{\partial E_{k}}\right| = |m_{\lambda}^{*} - m_{\lambda^{'}}^{*}|
\label{eqn:mass}
\end{equation}

\noindent
where $m_{\lambda}^{*}$ ($m_{\lambda^{'}}^{*}$) denotes the effective mass of path $\lambda$ ($\lambda^{'}$)~\cite{KartsovnikPhysRevLett.77.2530,HarrisonLaB6PhysRevLett.80.4498}. Note that it is the \textit{difference} in the effective masses of the two interfering paths that determines the apparent effective mass of QIOs -- thus enabling QIOs to be observed to much higher temperatures than QOs from the dHvA- and SdH-effects~\cite{harrison1996,KartsovnikPhysRevLett.77.2530,HarrisonLaB6PhysRevLett.80.4498,singletonrepprogphys2000}.

Figure~\ref{fig:wiggle3} shows that the $\upgamma$ and $\updelta$ frequencies in the QIO spectra of UTe$_2$ exhibit apparent effective masses ($\approx$~5~$m_{\text{e}}$) almost an order of magnitude lower than the quasiparticle effective masses reported for dHvA QOs ($\sim$~40~$m_{\text{e}}$)~\cite{AokidHvA_UTe2-2022,Eaton2024}, showing that the subtraction of masses between the two trajectories in Eqn.~\ref{eqn:mass} has almost cancelled out. By contrast, the $\upalpha$ and $\upbeta$ frequencies are much heavier with masses in the region of 20--35~$m_e$ (Fig. \ref{fig:wiggle3} and Fig. S3 in the Supplementary Materials). This implies that these MB networks span FS sections with a highly anisotropic distribution of the Fermi velocity, $v_{\text{F}}$. This is consistent with several experimental \cite{WrayARPES_PhysRevLett.124.076401, AokidHvA_UTe2-2022, Eaton2024} and theoretical \cite{PhysRevB.103.094504, DandD} studies that indicate the hybridization between U $f$-electrons with the U $d$-bands and Te $p$-bands, which provides the dominant contribution to the Q2D FS, can result in significant variations in the effective quasiparticle masses at points around the cylindrical sheets. We note that our uncertainty in $m^{*}_{\upalpha}$ and $m^{*}_{\upbeta}$ is considerably larger than for $m^{*}_{\upgamma}$ and $m^{*}_{\updelta}$ due to these frequencies only being observable near the base temperature of the $^{3}$He cryostat used for this measurement, with the uncertainty in temperature dominating the uncertainty in $m^{*}_{\upalpha,\upbeta}$. Further measurements in the experimentally challenging temperature--field regime of $\leq$~200~mK and $\geq$~40~T are required to carefully probe the anisotropy of $v_{\text{F}}$ around the FS of UTe$_2$, and thus to better understand the hybridization of the $f$, $d$ and $p$ bands.

In principle, an infinite number of MB networks could give rise to QIOs. Thus, it is expected that orbits of the type $\mathcal{A_\text{ACDEA}} - \mathcal{A}_\text{ABA} = (2\mathcal{A}_{h^{+}} + \mathcal{A}_\upbeta) - \mathcal{A}_{h^{+}} = \mathcal{A_\upbeta} + \mathcal{A}_{h^{+}}$ should occur. However, the effective mass associated with these orbits would be greater than the masses of the hole and electron orbits from which they arise. If in the most simple case we assume that the breakdown orbits of type $\mathcal{A_\text{ACDEA}}$ have masses of $2m^{*}_{h^{+}/e^{-}} + \epsilon_{m}$, where $\epsilon_{m}$ is a small difference to account for the fact that quasiparticles are in fact not traversing full FS sheets, then by Eqn.~\ref{eqn:mass} these breakdown orbits need to interfere with two full FS sheet orbits to produce oscillations of $m^{*}=\epsilon_{m}$. By comparison, orbits of the type $\mathcal{A_\text{ACDEA}} - \mathcal{A}_\text{ABA}$ would instead have masses of $m^{*}= m^{*}_{h^{+}/e^{-}} + \epsilon_{m}$ and as such would be too heavy to observe at $^{3}$He temperatures.

Figure~\ref{fig:wiggle4} shows the evolution of QIO frequency with magnetic field tilt angle, and compares with the prediction from our Q2D FS model (in panel c). Although this is only a crude approximation of the expected QIO frequency profile, we find remarkably good agreement between our FS model adapted from ref.~\cite{Eaton2024} and the QIOs we observe in TDO measurements. This result gives strong confidence that the FS of UTe$_2$ is very well described by our Q2D model.

Our discussion so far has focussed on field aligned coaxially to the FS cylinders (along $c$), and at inclination angles close to $c$. Figure~\ref{fig:wiggle5} shows that for field oriented along the $a$-axis, two additional frequencies $f_\upzeta$ = 220~T and $f_\upeta$ = 4.5~kT are observed. Again, the enclosed areas of these MB networks correspond very well to our Q2D FS model (Fig.~\ref{fig:wiggle5}f). The low frequency $\upzeta$ oscillations for field along $a$ are of considerable amplitude and are clearly observable in the raw TDO signal without background subtraction (Fig.~\ref{fig:wiggle5}a). Along the $a$-axis $\upzeta$ again corresponds to a QIO, whereas $\upeta$ is consistent with a conventional MB orbit, which may explain its small amplitude as well as its observation for $B$ only directly along the $a$-axis.

We note that a similar study of oscillations in the TDO signal of UTe$_2$ at high fields was recently reported~\cite{broyles2023revealing}. For $H \parallel a$ ref.~\cite{broyles2023revealing} reports an oscillatory frequency of 223~T, in very good agreement with the 220~T~$\upzeta$ orbit we observe at this field orientation (Fig.~\ref{fig:wiggle5}). However, rather than being of a QI origin, the authors of ref.~\cite{broyles2023revealing} interpreted the observed oscillatory waveform to comprise QOs from the SdH-effect caused by the presence of light 3D FS pocket(s). The distinction between Q2D and 3D FS dimensionality in the case of UTe$_2$ is important, as any 3D pockets could have significant implications regarding the topological properties of the putatively spin-triplet superconductivity~\cite{Korean_UTe2_topology_arxiv,Sato-Ando-topSCs_2017}.

However, in our measurements we do \textbf{not} observe any indication of the presence of a 3D FS pocket. Fig.~\ref{fig:wiggle5}b shows the evolution of $\Delta f_{\text{TDO}}$ as the field is tilted away from $a$ towards $c$. For magnetic field oriented along the $a$-axis we observe low frequency large amplitude oscillations, in good agreement with the raw data presented in ref.~\cite{broyles2023revealing}. A large oscillatory component is still visible 9$\degree$ away from $a$; however, after a rotation of 20$\degree$ (to $\theta_c = 70\degree$) no oscillations are observed within the resolution of the measurement. This is inconsistent with this frequency branch coming from SdH-effect QOs due to a 3D pocket; however, this behavior is consistent with a QI interpretation of the oscillatory origin, as the $\upzeta$ trajectory is only possible for $B$ close to $a$. In the Supplementary Materials we show a similar evolution for rotating away from $a$ towards $b$. Furthermore, no slow oscillations at these tilt angles have been reported in prior dHvA measurements by the field modulation~\cite{AokidHvA_UTe2-2022} or torque magnetometry~\cite{Eaton2024} techniques -- they appear only to be observed in the electrical conductivity, again consistent with a QI origin.

The stark difference in the effective masses of the $\upalpha, \upbeta$ and $\upgamma, \updelta$ components implies a strong anisotropy of $v_{\text{F}}(\textbf{k})$. In our recent study of dHvA QOs in UTe$_2$ we observed two-fold effective mass variations along the measured frequency branches under rotation away from the $c$-axis~\cite{Eaton2024}. In order to attain such a variation, this implies a significant anisotropy of $v_{\text{F}}(k_{z})$, which in turn could account for the large difference in effective masses of the QIOs. Such a variation in effective mass likely stems from substantial hybridization between U $d$-bands and Te $p$-bands, which are the main contributors to the Q2D FS sheets~\cite{DandD}, and an $f$-electron band sitting just above the Fermi level. This band has been detected in ARPES measurements, in which a significant spectral weight was observed at the Z-point of the BZ~\cite{WrayARPES_PhysRevLett.124.076401}. Models of UTe$_2$ that include the presence of such a band~\cite{PhysRevB.103.094504, DandD} show that the effect of the U $f$-electrons hybridizing with U $d$-bands is to compress them in energy, effectively increasing their band mass~\footnote{A similar effect, albeit less pronounced, may also be relevant for the Te $p$-band.}. It is therefore likely that $v_{\text{F}}$ is lowest (and thus $m^{*}$ is highest) at the regions of the FS cylinders that are closest to the Z point, as here the spectral contribution of the $f$-electrons is largest and thus the hybridization with them will be the greatest.


In summary, we measured the contactless resistivity of UTe$_2$ to high applied magnetic field strengths. We observed oscillatory components that are well explained by quantum interference between semiclassical quasiparticle trajectories spanning magnetic breakdown networks. We find that the quantum interference frequencies correspond very well to a quasi-2D model of the UTe$_2$ Fermi surface. Our observations give no indication of the presence of any 3D Fermi surface pockets in this material.

\vspace{-5mm}
\begin{acknowledgments}\vspace{-5mm}
We are grateful to N.R. Cooper, D.V. Chichinadze, D.~Shaffer, A.J. Hickey, H. Liu, P. Coleman, J. Chen, C.K. de Podesta, O.P. Squire, T. Helm, and especially A.F. Bangura for stimulating discussions. We thank T.J. Brumm and S.T. Hannahs for technical advice and assistance. This project was supported by the EPSRC of the UK (grants EP/X011992/1 \& EP/R513180/1). A portion of this work was performed at the National High Magnetic Field Laboratory, which is supported by National Science Foundation Cooperative Agreement No. DMR-2128556 and the State of Florida. We acknowledge support of the HLD at HZDR, a member of the European Magnetic Field Laboratory (EMFL). The EMFL also supported dual-access to facilities at MGML, Charles University, Prague, under the European Union's Horizon 2020 research and innovation programme through the ISABEL project (No. 871106). Crystal growth and characterization were performed in MGML (mgml.eu), which is supported within the program of Czech Research Infrastructures (project no. LM2023065). We acknowledge financial support by the Czech Science Foundation (GACR), project No. 22-22322S. T.I.W. and A.G.E. acknowledge support from QuantEmX grants from ICAM and the Gordon and Betty Moore Foundation through Grants GBMF5305 \& GBMF9616. A.G.E. acknowledges support from the Henry Royce Institute for Advanced Materials through the Equipment Access Scheme enabling access to the Advanced Materials Characterisation Suite at Cambridge, grants EP/P024947/1, EP/M000524/1 \& EP/R00661X/1; and from Sidney Sussex College (University of Cambridge).
\end{acknowledgments}

\bibliography{UTe2-QI}

\end{document}